# Anchoring Historical Sequences Using a New Source of Astro-chronological Tie-points

Michael W. Dee[1] and Benjamin J. S. Pope[2]

[1]University of Oxford – RLAHA, South Parks Road, Oxford OX1 3QY, United Kingdom.
[2]University of Oxford – Physics, Denys Wilkinson Building, Keble Road, Oxford, OX1 3RH, United Kingdom.

**Abstract**
**The discovery of past spikes in atmospheric radiocarbon activity, caused by major solar energetic particle events, has opened up new possibilities for high-precision chronometry. The two spikes, or Miyake Events, have now been identified in tree-rings from around the world that grew in the years 775 and 994 CE, exactly. Furthermore, all other plant material that grew in these years would also have incorporated the anomalously high concentrations of radiocarbon. Crucially, some plant-based artefacts, such as papyrus documents, timber beams, and linen garments, can also be allocated to specific positions within long, currently unfixed, historical sequences. Thus, Miyake Events represent a new source of tie-points that could provide the means for anchoring early chronologies to the absolute time-scale. Here we explore this possibility, outlining the most expeditious approaches, the current challenges and obstacles, and how they might best be overcome.**

**Introduction**
The annual history of Western civilisation only extends back as far as 763 BCE (Rawlinson 1867; Kitchen 1991), even though state-level societies emerged several millennia before this time. Similarly, Chinese history is only widely agreed from 841 BCE (Watson 1961). For earlier periods, historians and archaeologists use king-lists, stratigraphy, artefact continua, and other relative sequences to order events. The timelines that result are called 'floating' chronologies because they are fragmentary and unfixed on the absolute (BCE/CE) time-scale. Precise attribution of ancient astronomical observations can help anchor floating chronologies. For example, the aforementioned 763 BCE date relates to a solar eclipse recorded during the 9th year of Ashur-Dan III of Assyria (Rawlinson 1867). However, most early records of the night sky are either too ambiguous or, perversely, too common to pinpoint such chronologies. Some assistance is provided by the chronometric techniques, especially radiocarbon ($^{14}C$) dating, but even here estimates are only precise to within 200 – 300 calendar years. Bayesian modelling, which allows $^{14}C$ and relative data to be combined, has enabled some historical events to be constrained to within decades – but only in exceptional circumstances (see Bronk Ramsey *et al.* 2010; Dee *et al.* 2013). In any case, decadal resolution is still of limited value for the analysis of human societies. In short, the earliest periods of civilisation currently rest on insecure foundations, any comparisons between such societies should be considered provisional, and any alignments between cultural and climatic records at these times considered even more tenuous.

In this paper, we propose a new strategy for anchoring the floating chronologies of early history to the absolute (BCE/CE) time-scale, thereby extending the exact history of civilisation back several millennia. The method is also based on rare astronomical events, but employs the cosmogenic isotope $^{14}$C as the observer, and tree-rings as the natural almanac. In this study, our objectives are to outline the rationale and potential of the dating method, to define the remaining challenges, and to describe how they might best be overcome. Pragmatic considerations such as measurement costs and sampling permissions are not discussed.

**Miyake Events**
Radiocarbon is produced in the upper atmosphere primarily by the capture of thermalized neutrons by nitrogen (Libby *et al.* 1949; Beer *et al.* 2012; Kovaltsov *et al.* 2012). The neutrons are almost entirely liberated by cosmic ray spallation, although photoneutron reactions from γ-ray strikes are also a possible origin (Povinec & Tokar 1979; Pavlov *et al.* 2013; Carlson *et al.* 2010). The radioisotope rapidly becomes oxidised to $^{14}CO_2$, and evenly distributed around the globe. $^{14}$C is incorporated into plant tissue by way of photosynthesis and is thereafter transmitted up the food chain. When an organism dies, however, the $^{14}C/^{12}C$ ratio of its tissue declines exponentially due to beta decay. Thus, measurement of this ratio provides an estimate of the time elapsed since the organism was alive (Libby *et al.* 1949; Arnold & Libby 1949). Very soon after the invention of the $^{14}$C dating method, however, it became apparent that the data being generated needed to be corrected for past changes in the atmospheric activity of the radioisotope (Suess 1955, Tans *et al*. 1979). This process is called $^{14}$C calibration. In essence, it involves comparing the $^{14}$C measurement obtained on the sample with a reference curve of $^{14}$C measurements made on samples of independently established age (through dendrochronology or varve counting, for example). The Northern Hemisphere terrestrial calibration curve is known as IntCal13 (Reimer *et al.* 2013). Because of the propagation of measurement uncertainty in creating the $^{14}$C calibration record, and because the record fluctuates or even remains flat with time, individual sample measurements usually generate calendar date ranges of around 100-200 years.

The observed variation in atmospheric activity is largely a product of $^{12}CO_2$ emissions from volcanic and oceanic activity, but is also influenced by changing rates of cosmic flux (see Bronk Ramsey 2008). Interannual fluctuations were originally thought to be negligible (< 1-2 ‰, or < 8-16 $^{14}$C years); however, this assumption was radically undermined by Miyake *et al.* (2012), who published a major change in atmospheric concentrations (~ 12 ‰ or 100 $^{14}$C years) between the years 774–775 CE. The following year, the same team published a second anomaly, this time between 993 – 994 CE, similar in profile and magnitude to the first (~ 9 ‰, Miyake *et al.* 2013).

These rapid rises in atmospheric $^{14}$C concentration (henceforth, *Miyake Events*) inhere a combination of features that make them unique for high-precision chronometry. Firstly, they are easily distinguishable from the normal equable patterning of $^{14}$C in the atmosphere. In fact, the first year of a Miyake Event takes the form of an almost vertical spike in atmospheric activity (see Figure 1). Secondly, the exact calendar year in which a Miyake Event occurred is easily ascertainable, because tree-rings retain the $^{14}$C signal from the year in which they grew, and dendrochronological archives exist in which the growth year of every tree-ring is exactly known. Finally, Miyake Events are precisely synchronous and of comparable magnitude all around the Earth (Figure 1). The spike in 775 CE, for example, has already been found in tree-rings from Germany (Usoskin *et al.* 2013), Russia and the USA (Jull *et al.* 2013) and New Zealand (Güttler *et al.* 2015). Crucially, the enriched concentrations will also have been absorbed by all other growing plants at the time, including those latterly fashioned into cultural items.

Miyake Events are also exceptional because they denote significant increases in $^{14}$C. Decreases due to the natural emission of fossil $CO_2$ are common but, prior to the Nuclear Age, no known terrestrial process could have been responsible for such sudden and globally observed enrichments. Primarily on this basis, it was determined that the spikes must have been caused by pulses of radiation from space. The sun, at first, was not thought capable of emitting radiation of the required magnitude (Miyake *et al.* 2012; Pavlov *et al.* 2013; Hambaryan and Neuheusser 2013). However, the consensus is now that intense Solar Energetic Particle (SEP) events are indeed responsible (Melott and Thomas 2012; Usoskin *et al.* 2013; Guttler *et al.* 2015; Mekhaldi *et al.* 2015). SEPs can be the result of extreme solar flares or Interplanetary Coronal Mass Ejections (ICMEs); although the pathways by which they increase spallation in the atmosphere and hence intensify $^{14}$C production are still being resolved (Usoskin & Kovaltsov 2012; Cliver *et al.* 2014).

**Use as Exact Time-Markers**
Miyake Events provide an opportunity for the precision of dendrochronology to be combined with the versatility of $^{14}$C dating. In a sense, they allow for the exportation of tree-ring dates to a variety of other materials. The dating process requires two main steps. Firstly, more events must be uncovered in the dendrochronological archives, and thus dated to the exact calendar year. In two cases, the 775 and 994 CE Events, this has already been done. However, it is currently unclear how regular the events are, in either magnitude or occurrence frequency. This paucity of information is a product of the misconception that annual $^{14}$C variation was negligible and because $^{14}$C measurements are comparatively expensive to obtain. In the construction of the IntCal13 calibration record, Reimer *et al.* 2013), for example, measurements were mostly made on decadal blocks of tree-rings, a technique that completely denuded the annual substructure of the record.

The recent attribution of the single-year spikes to SEP events greatly improved the probability that many more will be discovered in the Holocene. This can be concluded because the activity of the sun is marked by its cyclicity, and because the two events discovered thus far are comparatively close together. The galactic and extra-galactic causes originally mooted, on the other hand, are now much less likely as they are thought to occur on multimillennial time-scales (Pavlov *et al.* 2013; Cliver *et al.* 2014, Melott *et al.* 2015). Potential strategies for uncovering further Miyake Events are addressed further below.

The second step of the dating process requires $^{14}$C measurements to be made on samples from plant-based artefacts that come from successive calendar years. If one of the samples overlies a known spike, the entire sequence can be automatically allocated to exactly one point in time. The simplest working example involves applying the technique to another piece of wood, such as a beam from a timber-framed house. It would first be necessary to determine whether the object might date near an anomaly. This could be established by making a few $^{14}$C dates in the normal fashion, or indeed it may be known *a priori* on the basis of the archaeo-historical context. Subsequently, annual or biennial tree-ring samples are measured until the anomaly is found. From this datum, the number of rings outward to the bark edge would just need to be counted to obtain the exact calendar year for the felling of the tree. This approach has already proven successful. Wacker *et al.* (2014) used the spike in 775 CE to date a chapel in Val Müstair, Switzerland to 786 CE. The universality of the method becomes apparent when one realises that the same signal will be present in every wooden building in the world constructed at this time. But the greatest potential of the approach pertains to the case where the object itself also has a defined position within an independent chronology. For ancient history, this means artefacts attributed to specific years of floating chronologies. The approach is exactly analogous to the use of Miyake Events to anchor the Greenland ice-core chronologies, which has also just been accomplished (Sigl *et al.* 2015). Indeed, the tie-points are being sought in other environmental archives as well, such as coral sequences (Lui *et al.* 2014, Ding *et al.* 2015). However, in many respects, the archaeological scope and benefits are potentially even more profound, as exemplified by the following two key chronologies.

*The Mesoamerican Long Count (MLC)*
The MLC is a day numbering system that was used in pre-Columbian America, underpinning the chronology of the Classic Maya and other contemporary civilisations. It spans at least a full millennium although it not known whether any inconsistencies exist in the sequence over this time (Thompson 1937; Thompson 1950). The MLC has never been precisely connected to the absolute (CE/BCE) calendar. The most widely accepted alignment is called the Goodman-Martínez-Thompson (GMT) correlation but others, which vary by as much as centuries, are also in use (Thompson 1937; Kennett 2013). As a result, the entire MLC can be regarded as a long floating chronology. In fact, the earliest truly fixable date in the Americas is still taken to be the arrival of Columbus in 1492 CE.

Some wooden lintels from Maya temples are inscribed with MLC dates. Recently, a suite of $^{14}$C measurements were made on one such artefact from the site of Tikal, and subsequently incorporated in a Bayesian chronological model. The felling date produced of 658-696 (CE, 95% probability) was consistent with the GMT correlation (Kennett *et al.* 2013). However, the discovery of a Miyake Event in an artefact like this would potentially secure it to the exact year, concomitantly anchoring the entire Mesoamerican chronology. Care would have to be taken to distinguish annual rings, and isotopic methods may prove best for this (Cernusak & English 2015), but the opportunity is striking as the lintels of

Tikal are tantalising close to the 775 Event that has already been used to date the chapel at Val Müstair.

*The Egyptian Chronology*
The precision and longevity of the Egyptian chronology is unrivalled in the ancient world. It comprises a sequence of kings and queens and the lengths of their reigns. The most important sections are those that 'float' over the 3rd and 2nd millennia BCE, as few other historical timelines are available for this period. The chronology is also central to the rise of Western civilisation, as a number of contemporaneous societies are effectively dated by tracing material synchronisms back to Egypt (Kitchen 1991; Hornung *et al.* 2006). The historical record comprises periods where long contiguous sections are common, called the Old, Middle and New Kingdoms, interspersed by Intermediate Periods of unknown duration. One of the longest continuous sections (*ca.* 200 years) comes from the Middle Kingdom. It is currently situated in absolute time using one text foretelling the heliacal rising of Sirius from the mid-19th century BCE (Parker 1950). But this tie-point is much disputed and estimates of the true position of the sequence still vary by more than 50 years (Kitchen 1991).

One of the advantages of archaeological sequences is that the group of samples might not even need to come from the same object. It would suffice to obtain a series of annual plant materials that are explicitly from successive calendar years. So, for example, a series of individual papyri from consecutive regnal years would act in the same way as a series of individual tree-rings; although, the possibility of palimpsests would also have to be taken into account. Assuming the papyri were all used soon after they were harvested, it also follows that not every year of the sequence would have to be covered. An anomaly as pronounced as the two discovered so far would be evident in biennial or even triennial samples (see Figure 2). Further still, the artefact being measured would not even need to be directly labelled with the cultural date. For example, cedar beams that contain tens of annual rings are available from the Bent Pyramid of king Sneferu. There is strong archaeological evidence that cedar obtained from Lebanon, and used in the Bent Pyramid, was obtained during the earliest years of Sneferu's reign (see Verner 2001: 177). Indeed, it is likely that his accession date was no more than ten years before this event. Using this assumption, and the hypothetical discovery of a Miyake Event somewhere the cedar tree-rings, the decade within which his accession took place could be secured. Moreover, even just fixing this one decade would radically improve the chronology of the Eastern Mediterranean during this important time period. This effect is illustrated in Figure 3 using simulations in the OxCal chronological modelling program (Bronk Ramsey 1995).

Although the most appropriate samples for the method are annually resolved plant archives, spikes may also be detectable in less precisely defined contexts. For example, if anomalies in $^{14}C$ content are identified between plants within the same structure, such as reed matting in a mud-brick building, or harvest grains in the same silo, it may be possible to ascribe an exact calendar year to the feature in question. However, care would have to be taken to exclude the possibility that the spike did not just arise from intrusive younger material (and therefore richer in $^{14}$C). A further goal might also be to identify a $^{14}$C spike in animal remains. Because of the complexity of animal metabolism and tissue turnover, it is only likely that this will be realised if a specific compound of plant origin, which is rapidly directed into short-lived tissue such as skin or hair, can be isolated and measured.

**Uncovering New Events**

It is important to stress that this method could immediately be rolled out for any historical sequences that traverse the 775 and 993 CE anomalies. Indeed, chronologies that straddle both events offer the prospect of dual tie-points, and hence checks on the internal consistency of the record at hand. However, in order for the method to be extended to the earliest millennia of civilisation, new spikes need to be found. The two main strategies that could be deployed toward this goal are the interpolation of existing data sets and analysis of historical records.

*Gaussian Process Modelling*

Three cosmogenic isotope time series exist that should preserve information on the past occurrence Miyake Events: the $^{14}$C calibration curve (tree-ring) data, and the $^{10}$Be and $^{36}$Cl (ice-core) measurements (e.g.: Yiou *et al.* 1997; Baumgartner *et al.* 1998). This is because their production pathways and cross sections are similar, albeit not identical (Beer *et al.* 2012; Pavlov *et al.* 2013). In their current form, however, these data are too coarse to reveal the single-year spikes. As previously mentioned, the $^{14}$C data mostly represent averages over decades; and the ice-core results tend to be point estimates every 10 years (or more). It is not practicable to make the thousands of measurements necessary on known-age tree-rings to determine directly when Miyake Events occurred. Fortunately, Gaussian Process (GP) modelling is having considerable success at uncovering 'change points' hidden within time series analogous to cosmogenic isotope records (Roberts *et al.* 2012). A Gaussian Process provides a probability distribution over functions as an extension of how a multivariate Gaussian is a distribution over vectors: any finite collection of samples from the (infinite-dimensional) function is itself distributed as a multivariate Gaussian, with the covariance matrix determined by a 'kernel', which is a function of the input variable (for example time); different kernels encode information about the sorts of variation that are expected in functions drawn from the GP (Rasmussen & Williams, 2006).

In specific terms, there are two main ways in which GPs could help to determine when sudden increases in $^{14}$C occurred. Firstly, they have the capacity to extract information on gradient from even a sparse or noisy function, which is the prime conveyor of information regarding dislocations and abrupt changes in the data. Secondly, they can be engineered around 'change point kernels,' which explicitly address the question of whether a change has taken place in the data. In addition to this, multi-input multi-output GP approaches can simultaneously combine information from radiocarbon and other sources, such as ice-core isotope samples. By applying these analytical strategies to the three data sets, alongside more obvious patterning like possible periodicity, the years of highest probability should be disclosed.

To demonstrate the efficacy of this approach, we have run simulations using more elementary GP models of the IntCal13 time series (Figure 4) across the Holocene. We intentionally restrict our time series to only decadal averages, even though at some times data are available with higher time resolution, in order to obtain a homogeneous dataset to demonstrate the technique; different noise processes at different resolutions would otherwise necessitate a more detailed GP model. Noting there are very-long-term (~ thousand year) and short-term (~ decades) variations, we choose a GP kernel consisting of a white noise kernel, plus two squared exponential kernels, which describe time series whose variations are smooth. One is initialized with a 20-year timescale, and the other with a 1000-year timescale. We then locally optimize these hyperparameters to obtain the best possible likelihood with respect to the time series. In future work, we may deliver more precise and accurate results using more sophisticated approaches: for example incorporating heterogeneous data obtained at different resolutions, where short timescales resolve the solar cycle; more appropriate kernels, such as Matérn kernels which allow for more roughness; multi-input multi-output models; and explicit change-point techniques.

Table 1 lists the most prominent changes in atmospheric $^{14}$C, as detected by the algorithm, both upwards and downwards. These are produced by ranking the largest positive and negative excursions from the GP predictive mean, divided by the GP uncertainty at each point; we display only the top seven such candidate years in each direction. It should be noted that these outputs are provisional, and also entirely dependent upon the accuracy of the raw measurements themselves. With several exceptions, these event candidates are rarely more than three sigma away from the mean, and therefore it is certainly not the case that each uptick will represent a Miyake Event: many will turn out to be 'false positives' and, where the data are too poor, spikes may have been missed. The 775 CE event is easily detected, and it is not even the most dramatic excursion in the data set over this time (see Figure 4). The known 994 CE event, however, is not so obvious. Nonetheless, by applying these approaches to all the cosmogenic isotope data sets, and uniting the outputs with other sources of information (see historical observations below), the process of uncovering past events should be accelerated.

As previously discussed, decreases in $^{14}$C can be caused by any number of volcanic, oceanic or potentially even burning events. The signals from such events, however, can be delayed, or attenuated over several years, so attributing them to specific occurrences can be problematic. However, if a sharp decline is detected, better still one adjacent to a sharp increase, such as is apparent for the years 676 and 656 BCE, this may form a unique pattern in the annual $^{14}$C record that would make for an even more reliable time marker.

| Prominent Increase Years | Prominent Decrease Years |
| --- | --- |
| 3886 BCE | 4126 BCE |
| 3076 BCE | 3876 BCE |
| 2387 BCE | 1327 BCE |
| 1677 BCE | 676 BCE |
| 656 BCE | 456 BCE |
| 544 CE | 794 CE |
| 774 CE | 1044 CE |

**Table 1. The years identified by the Gaussian Process model as most likely to represent when sudden changes in the atmospheric radiocarbon concentration occurred, including both enrichments and depletions.**

*Historical Observations of Aurorae*
Incident SEPEs disrupt the Earth's magnetic field directing streams of charged particles into the Polar Regions enlarging the red, green and blue aurorae (Cliver & Svalgaard 2004). The Carrington Event was the strongest ICME witnessed in modern times and the disturbance caused to the Earth's magnetic field was so great that aurorae were observed as far south as Hawaii (Cliver & Svalgaard 2004). Many early societies were obsessive recorders of the night sky, and such records could be scoured for clues to dramatic aurorae events (Aaboe 2001). Any such attestations should be dealt with cautiously, in consultation with experts, but it is clear that mutually corroborating records are present in the historical record. For example, identical observations are found in records for geographically separate civilisations of East Asia (Willis & Stephenson 2000), and it has been postulated that 775 Event was alluded to in the *Anglo Saxon Chronicle* (Allen 2012).

**The Prospect of Exact Ancient History**
Miyake Events present a new paradigm for chronology. Because of their annual resolution and global occurrence, only a handful of tie-points would be required to revolutionise current understanding of early civilisation. Anchoring records in absolute time will expose ancient history to the level of scrutiny considered essential in modern history. Processes of cause and effect in human societies, even those of long-term consequence, commonly take place on annual or sub-annual time-scales. The inability of existing chronometric methods to unpick such processes is manifest. For example, if the $20^{th}$ century CE were visible at the resolution currently available for the $20^{th}$ century BCE, the two World Wars would currently be indistinguishable in time.

Furthermore, if more and more parochial records were secured by the annual tie-points, a chronological lattice would emerge within which the pattern of knowledge and technology flow could be readily observed. For example, if two adjacent civilisations with their own chronologies, such as Egypt and Assyria in the $2^{nd}$ millennium BCE, could be fixed to the same Miyake Event, new questions could be addressed about the exchanges between them. The passage of other important ideas, which may not leave directly datable evidence, such as mathematical, ideological or religious concepts, would also be traceable across space, as each culture would be secured to the same time frame.

**Conclusion**

Astronomical observations have long been essential to chronology. Single-year spikes in the Holocene $^{14}$C record present an entirely new type of astro-chronological tie-point. In this study, we have introduced both the theoretical and practical aspects of a employing these events for high-precision dating. It is envisaged that the method will ultimately provide important new insights into early civilisation.

**Acknowlegements**
The authors would like to thank Professor Suzanne Aigrain for her comments on the draft.

**Ethics statement**
This work did not involve any collection of human data.

**Data accessibility statement**
This work does not contain new experimental data.

**Competing interests statement**
We have no competing interests.

**Authors' contributions**
MD conceived the dating approach and drafted the paper. MD & BP collectively developed the ideas and their implications. BP wrote and defined the Gaussian Process analysis.

**Funding**
MD is supported by a Leverhulme Trust Early Career Fellowship (ECF-2012-123). This study built on an earlier investigation the authors conducted into supernovae and Miyake Events, supported by the Balliol Interdisciplinary Institute (BP as PI; MD as Co-I).

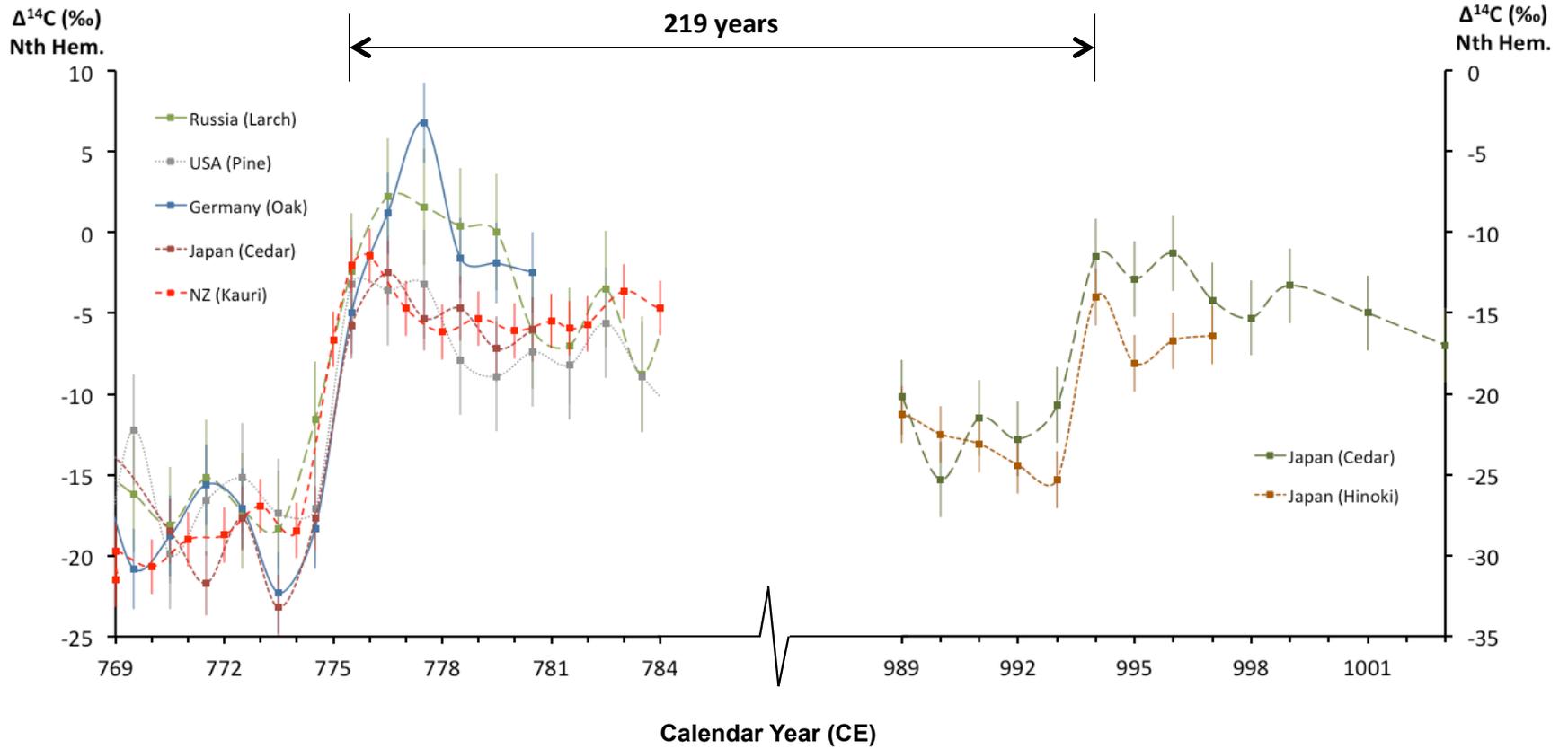

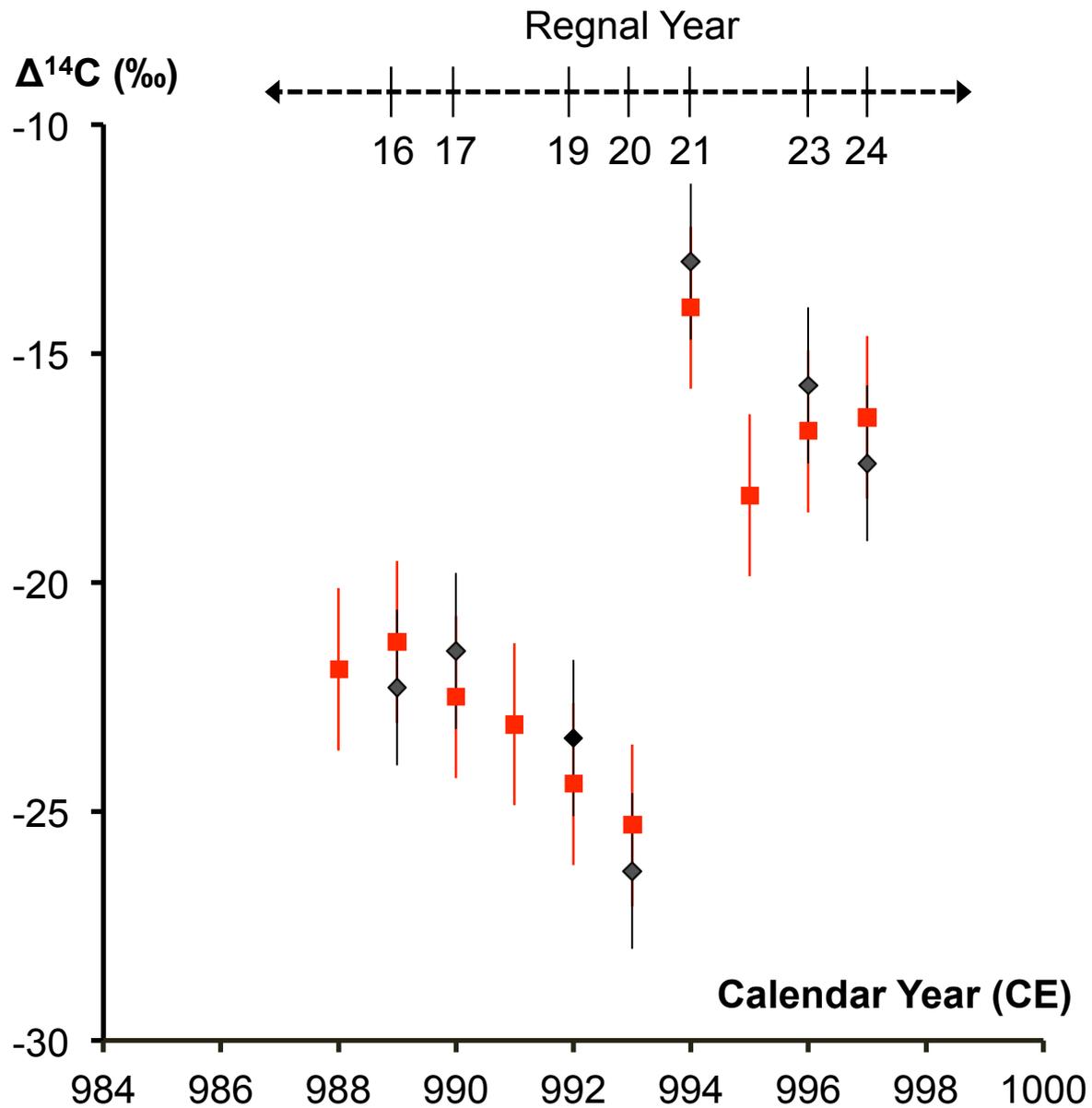

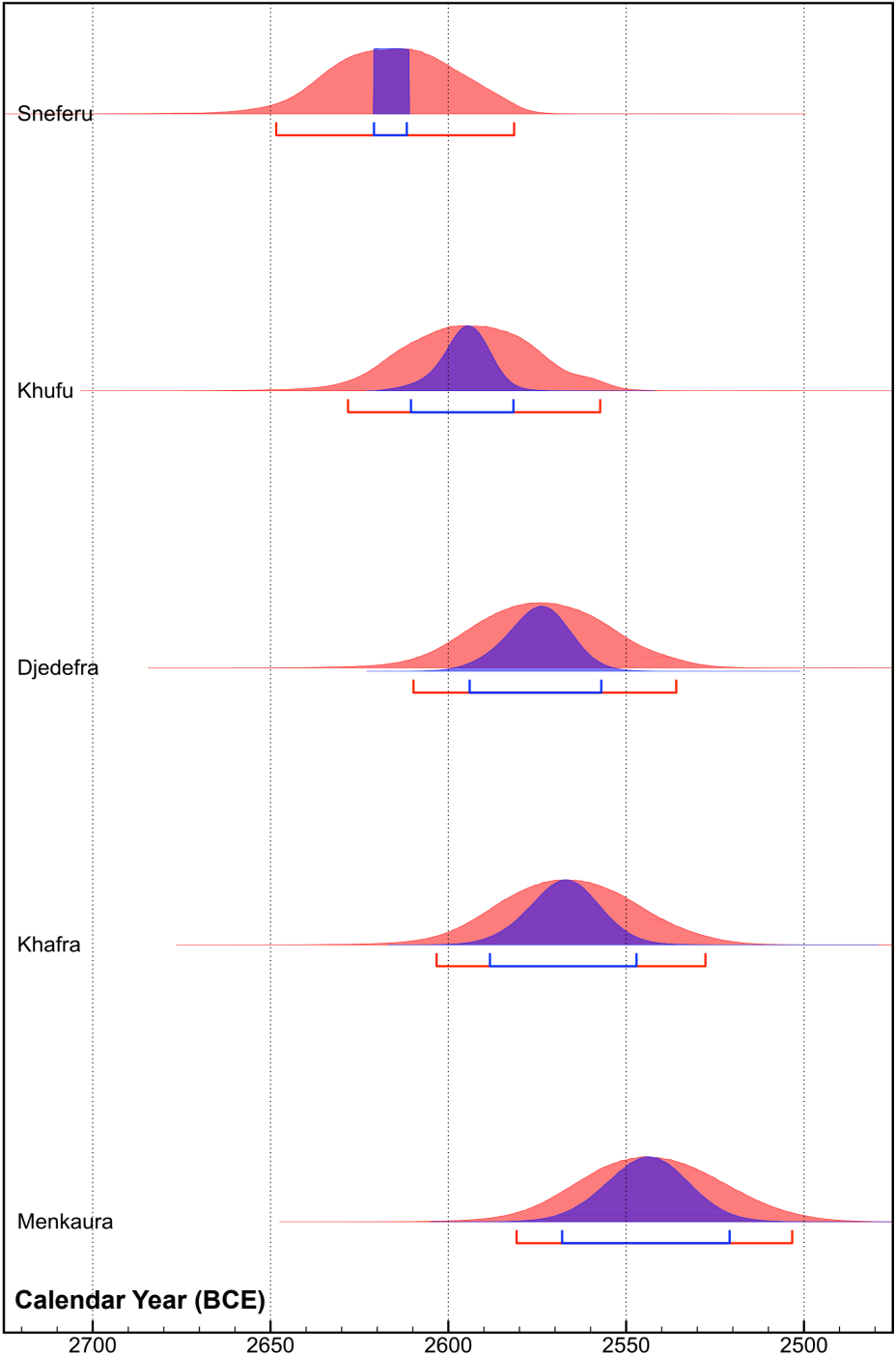
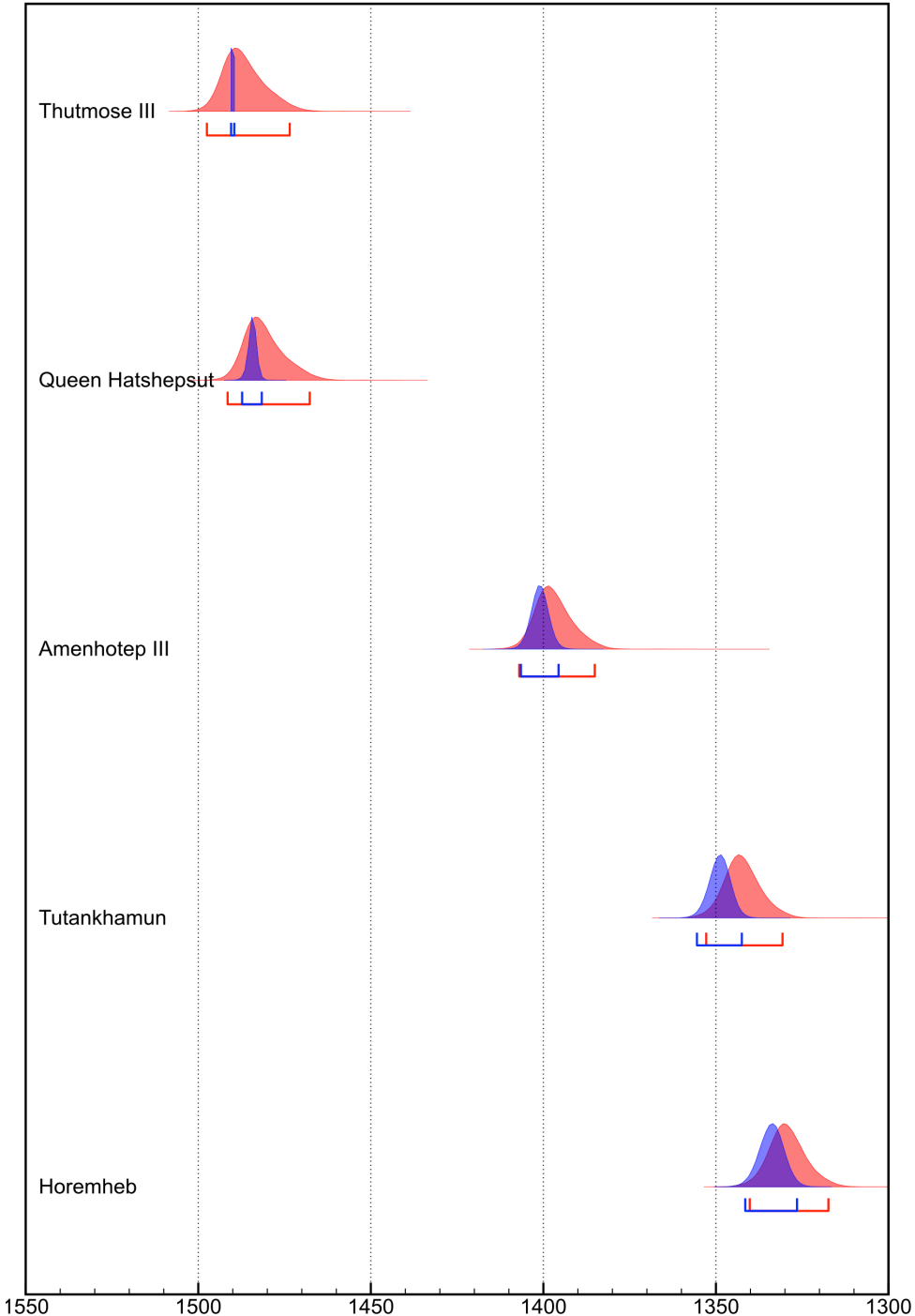

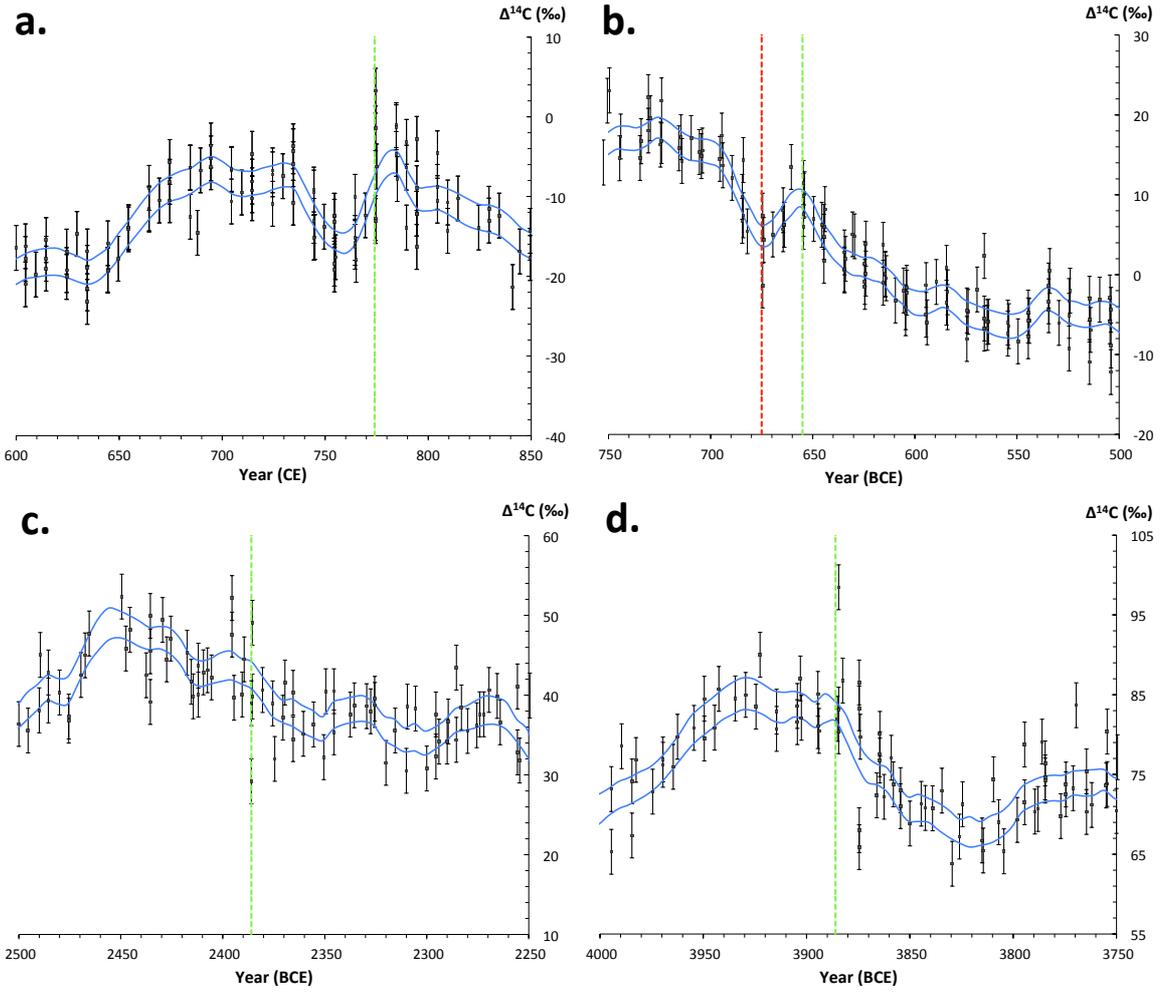

Figure 1: Time profiles of the measured Δ¹⁴C content in tree-rings from different tree species and different locations around the world. The spikes are unequivocal, and separated by exactly 219 calendar years. For ease of display, the data obtained on the kauri samples from New Zealand have all been elevated by 5‰, an amount which approximately corresponds to the offset between northern and southern hemisphere Δ¹⁴C concentrations.

Figure 2: A comparison between current 95% probability ranges (paler densities) and what could be achieved by the new dating approach (darker densities) for two important sections of the Egyptian chronology. By fixing the accession date of king Sneferu to within one decade (left-hand panel), the dates for a number of his successors are also significantly refined. If a single-year match, like that described in Figure 2, were achieved for items from the reign of Thutmose III, the uncertainty propagated forwards is solely a product of the historical knowledge of succeeding reign lengths. The OxCal model code for these simulations, which are modifications to Bronk Ramsey et al. (2010), is given in the electronic supplementary material.

Figure 3: A schematic of how the dating process would work for samples other than tree-rings. The squares represent the actual data for the 994 CE production spike from Miyake et al. (2014). The diamonds are hypothetical data points for various documents dated to a ruler's reign. To effect a match, it is not necessary for samples to be found for every regnal year.

Figure 4: Some of the key anomalies in past Δ14C values identified by the Gaussian Process analysis: the uplift between 774 and 775 CE (a); a rapid decrease closely followed by a rapid rise in the middle of the Iron Age plateau (b); a sudden uplift in the Early Bronze Age (c); and the most dramatic rise detected in the late Holocene (d).